\documentclass{article}        % Your input file must contain these two lines
\usepackage{epsfig,bm,float}
\usepackage{algorithm}
\usepackage[noend]{algpseudocode}
\usepackage{calc}
\usepackage{caption}
\usepackage{wrapfig}
\usepackage[table,xcdraw]{xcolor}
\usepackage{graphicx}
\usepackage{setspace}
\usepackage{amsmath}
\usepackage{amssymb} 
\usepackage{physics}
\usepackage{braket}
\usepackage{fullpage}
\usepackage{setspace}
\begin{document}               % plus the \end{document} command at the end.

\begin{center}
{\setstretch{1.1}
%\noindent {\Large \textbf{Unambiguous physical modeling from microwave spectroscopy}} \\
\noindent {\Large \textbf{Automated, context-free assignment of asymmetric rotor microwave spectra}} \\
\vspace{0.1in}
}
{Lia Yeh, Lincoln Satterthwaite, and David Patterson \\ davepatterson@ucsb.edu \\}
Broida Hall, University of California, Santa Barbara, Santa Barbara, CA 93106
\end{center}

\begin{abstract}
We present a new algorithm, Robust Automated Assignment of Rigid Rotors (RAARR), for assigning rotational spectra of asymmetric tops.  The RAARR algorithm can automatically assign experimental spectra under a broad range of conditions, including spectra comprised of multiple mixture components, in $\lesssim$ 100 seconds.  The RAARR algorithm exploits constraints placed by the conservation of energy to find sets of connected lines in an unassigned spectrum.  The highly constrained structure of these sets eliminates all but a handful of plausible assignments for a given set, greatly reducing the number of potential assignments that must be evaluated. 
We successfully apply our algorithm to automatically assign 15 experimental spectra, including 5 previously unassigned species, without prior estimation of molecular rotational constants.  In 9 of the 15 cases, the RAARR algorithm successfully assigns two or more mixture components.

%Microwave spectroscopy provides our most accurately known molecular structures.
% . Assignment of such spectra is a prerequisite for extracting meaningful information, such as molecular structures, from these spectra.

%Typically, such spectra are assigned by human experts identifying patterns withing the spectra, often using prior estimates of molecular constants obtained from quantum chemical calculation.  There has been significant recent progress in automated assignment of such spectra, including approaches XXX, to date no algorithm has demonstrated the robust ability to assign experimental spectra without prior knowledge of the compounds being analyzed. 

\end{abstract}
\doublespacing
\section{Background}

Microwave spectroscopy provides our most accurate measurements of molecular structures. Typical spectra recorded by modern instruments are comprised of thousands of lines, each corresponding to a specific rotational transition of a specific species.  Assigning the correct quantum numbers to lines in an observed spectrum is a prerequisite for extracting meaningful structural information about the molecule.  Today, such assignments are typically performed using a combination of quantum chemical simulation, which can in most cases predict rotational constants to within a few percent, and often laborious inspection of the observed spectrum. Assignment of complex spectra remains very much an art, and veteran spectroscopists employ diverse tricks as well as deep intuition to find requisite patterns amid congested spectra \cite{cooke2012decoding}.

Robust automatic fitting of rotational spectra or rovibronic spectra has been a longstanding goal of the spectroscopy community.  Attempts at fully automated algorithms include genetic algorithms \cite{hageman2000direct}, broad searches combined with quantum chemical calculation \cite{seifert2015autofit,shipman2019autofit}, assignment via nonlinear spectroscopy \cite{martin2016automated,fritz2018conformer}, and artificial neural networks \cite{zaleski2018automated}. 
{Colin Western's PGOPHER program \cite{western2017automatic} includes an implementation of the automated fitting routine described in reference \cite{seifert2015autofit}.}  {Of particular note are the genetic algorithms demonstrated by Meerts and Schmitt, which have successfully assigned rotationally resolved electronic spectra with no need for prior estimation of molecular constants \cite{leo2006application,schmitt2004determination}. These algorithms furthermore can be applied to a wide range of candidate Hamiltonians, in contrast to the rigid rotor Hamiltonian assumed in this work.  Automated, context-free assignment and structure determination from high-dimensional NMR data is now an integral part of modern NMR analysis \cite{moseley1999automated,evangelidis2018automated}}. 
In many cases approaches to spectral assignment leverage the ability of modern quantum chemical calculations to predict the structure of a compound, and thus the approximate rotational constants, from the elemental composition and connectivity of the compound, vastly reducing the size of the search space.  This approach is not applicable in the case where the identity of the analyte is unknown.  An algorithm which can assign spectra without the context of a prior prediction is therefore desirable. To our knowledge, no reliable algorithm has been demonstrated which can be used by a non-expert to assign the rotational spectra of mixtures of asymmetric tops without prior estimation of rotational constants.

Today, assignment is often done by experts using a combination of intuition and software, including Colin Western's PGOPHER program and Herb Pickett's SPFIT/SPCAT software suite \cite{PGOPHER,pickett}.  These tools can rapidly predict a spectrum from molecular constants, and can be used to find rotational constants given a correct set of observed frequencies and corresponding line assignments.  Our algorithm uses the SPFIT package to find rotational constants once it has identified a short list of plausible assignments.

 % We demonstrate here an algorithm which can robustly assign spectra of rigid asymmetric top molecules, without quantum chemical calculation or any prior knowledge of the chemical species in question.

  We present here an algorithm, Robust Automated Assignment of Rigid Rotors (RAARR), which can rapidly determine the rotational constants of multiple species within an unassigned spectrum under a broad range of conditions.  The RAARR algorithm is $\emph{context-free}$, meaning that no prior estimation of $A$, $B$, and $C$, or in fact, any information about the molecular species, is needed.  The RAARR algorithm is $\emph{consistent}$, using no pseudorandom number generation.  Lastly, the RAARR algorithm is \emph{fully automatic}, needing no user input other than the spectrum itself.
  
\section{Rotational spectra}
Many molecules are well approximated as rigid rotors; even flexible molecules typically fold into one of several rigid conformers at low temperature, with each conformer behaving separately as a rigid rotor or a slightly perturbed rigid rotor.  The rotational transition frequencies of a rigid rotor are determined by the rotational constants $A$, $B$, and $C$; these constants are related to the principal moments of inertia of the molecule $I_x$, $I_y$, and $I_z$ via
\begin{equation}
        A = \frac {h}{8 \pi I_x}\qquad
        B = \frac {h}{8 \pi I_y}\qquad
        C = \frac {h}{8 \pi I_z}
\end{equation}
  We confine ourselves here to the asymmetric top case, $A \neq B \neq C$, and the case where dipole moments $\mu_a$ and $\mu_b$ are non-zero. This is by far the most common case, and represents a ``typical'' molecule of low ($C_1$) symmetry.

\begin{figure}[!ht]
    \centering
    \includegraphics[width=5in]{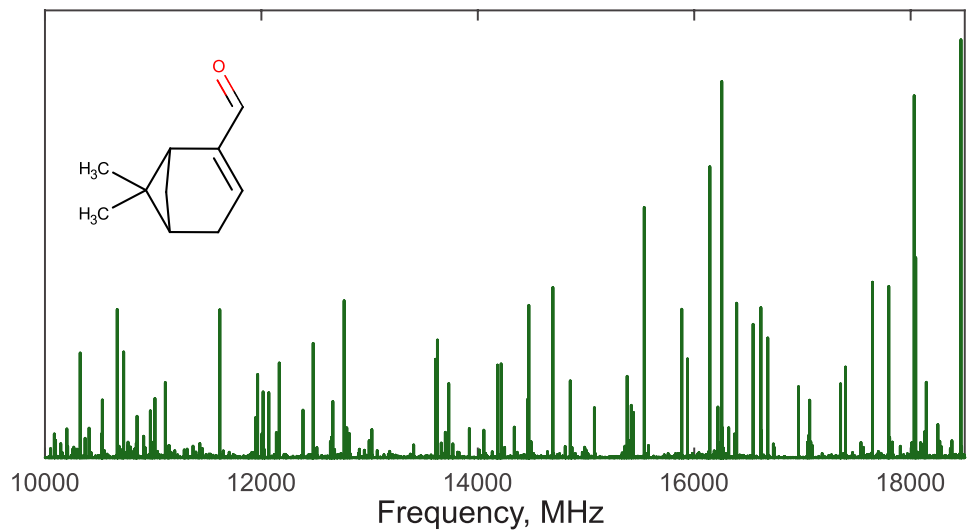}
    \caption{The microwave spectrum of myrtenal. \label{myrtenal_raw}}
    \end{figure}

  Predicting the microwave spectrum given $A$, $B$, and $C$ is straightforward, and can be done by several publicly available software packages, including PGOPHER by Colin Western and the SPFIT/SPCAT suite by Herb Pickett.  The reverse problem --- determining $A$, $B$, and $C$ from a microwave spectrum, is substantially more difficult.  Line assignments are non-obvious, and it is typically difficult to determine which lines in a complex spectrum belong to a common species.  A typical experimental spectrum is shown in Figure \ref{myrtenal_raw}.
  
   The microwave spectrum of a rigid rotor molecule is fully characterized by the rotational constants $A$, $B$, and $C$, and the corresponding electric dipole moment components $\mu_a$, $\mu_b$, and $\mu_c$.  Each transition's intensity is proportional to either $\mu_a$ if it is an $a$-type transition, $\mu_b$ if it is $b$-type, or $\mu_c$ if it is $c$-type.  The RAARR algorithm requires $\mu_a$ and $\mu_b$ to be non-zero; $\mu_c$ can be either zero or non-zero.
   In many spectrometers, including our own, line intensities are found experimentally to be unreliable, which essentially requires that any assignment algorithm be robust against unknown intensity fluctuations.  Our algorithm assigns the common case of ``nearly rigid'' molecules by the well established method of first approximating them as exactly rigid, and using the subsequent assignment to find centrifugal distortion  coefficients \cite{GordyCook}.
 
% \section{The RAARR algorithm}
% \begin{figure}[h]
%     \centering
%     \includegraphics[width=0.5\textwidth]{raarr_cfg.png}
%     \caption{A graphical control flow graph of RAARR.}
%     \label{fig:cfg}
% \end{figure}

The RAARR algorithm proceeds in the following steps, described below: 

\begin{enumerate}
    \item A simple peak-finding algorithm finds spectral lines from the original 1D spectrum, and all of these lines are marked as unassigned.
    
    \item The algorithm finds closely associated sets of unassigned lines, referred to in this paper as ``scaffolds,'' in the experimental spectrum. While many sets of associated lines exist within any spectrum, the algorithm searches only for the exact structure shown in Figures \ref{fig:patterns}D, \ref{fig:construction}C, and \ref{fig:scaffoldfig}B.
    The highly constrained structure of a scaffold makes it highly unlikely that non-associated lines will be falsely identified as a set. A typical scaffold in the molecule myrtenal is shown in Figures \ref{fig:patterns}D, \ref{fig:construction}C, and \ref{fig:scaffoldfig}B.
    
    \item The lines in each scaffold can be plausibly assigned in only a few ways.  Each of these assignments are attempted; those assignments which converge yield a candidate set of rotational constants $[A,B,C]$. In practice, a single set of constants $[A,B,C]$ is typically found for any given scaffold.
    
    \item A spectrum based on each candidate set of constants is calculated and compared against the original spectrum.  If a convincing match is found, the assignment list is incrementally extended to include predicted lines.  At this stage algorithm additionally fits the centrifugal distortion constants $\Delta_J, \Delta_{JK},\Delta_{K},\delta_J,$ and $\delta_K$.% and uses the fit values of these constants if they improve the predictive power of the fit.
    
    \item If a match is found, the lines that were correctly predicted in step 4 are marked as assigned, and the algorithm begins again at step 2 to search for additional species. 
\end{enumerate}
 
\subsection{Pattern finding}
\label{sec:pattern}
The scaffolds found in step 2 are constrained to simultaneously follow two patterns, ``4-loops'' and ``series,'' as described below.  

\begin{figure}[!ht]
    \centering
    \includegraphics[width=5in]{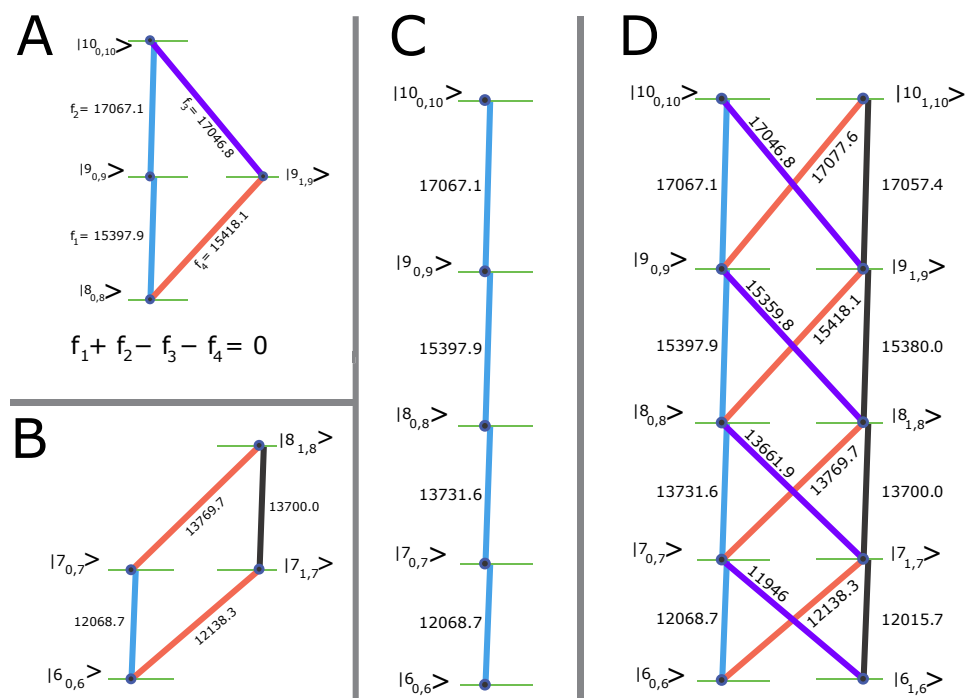}
    \caption{Loops and series are combined a larger pattern we call a ``scaffold''.  A, B: 4-loops, comprised of four lines which are constrained by conservation of energy to sum to zero. Each 4-loop shown here has two $a$-type lines (blue and/or black) and two $b$-type lines (red and/or purple).  C: A series, which is comprised of nearly evenly spaced lines.  D: A scaffold, which is comprised of four series (two $a$-type, two $b$-type), and many 4-loops.  This structure is highly constrained, and can be recognized in a spectrum prior to assignment.  The RAARR algorithm works by finding this specific, interwoven structure amid a complex spectrum.}
    \label{fig:patterns}
\end{figure}

\begin{figure}[!ht]
    \centering
    \includegraphics[width=6.0in]{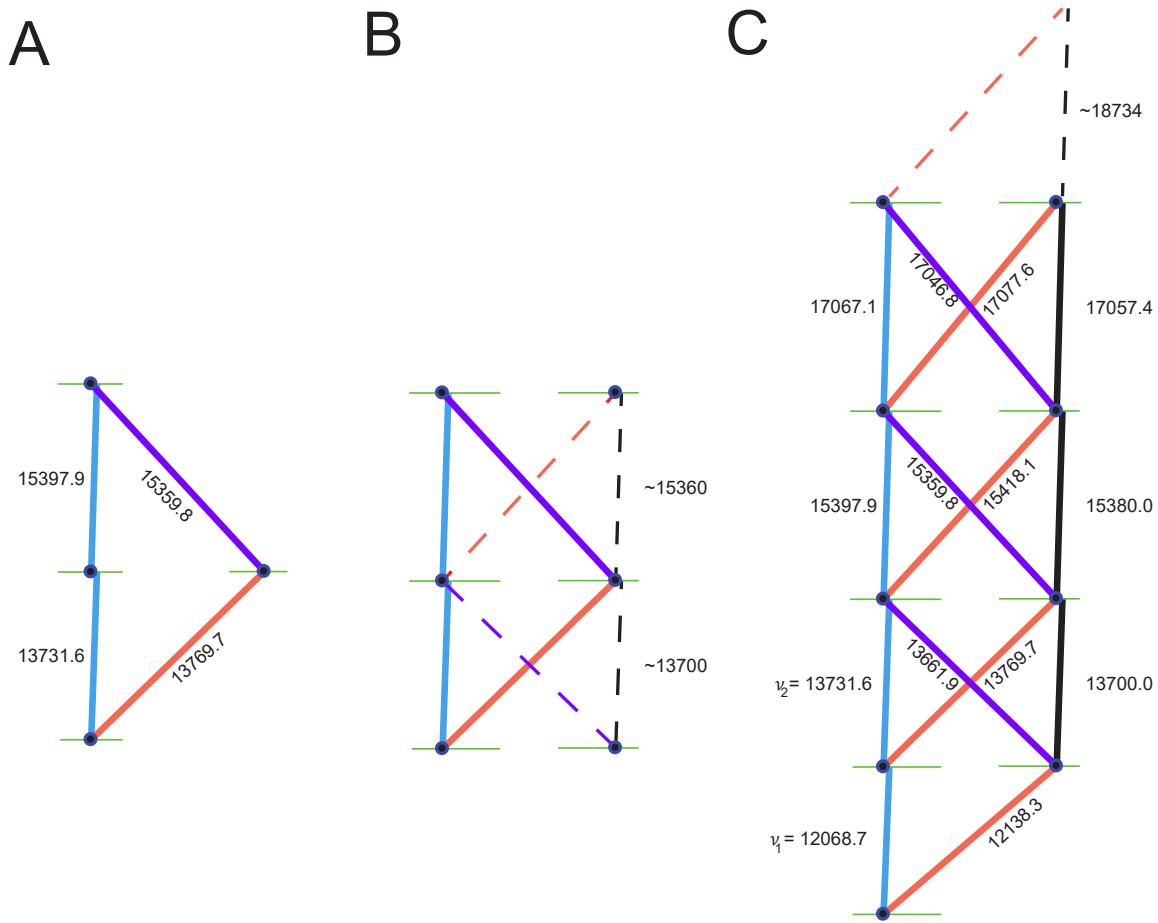}
    \caption{Scaffold construction. A: The original 4-loop is found by brute-force search, constrained that at least one line is in the strongest 40 lines in the spectrum. Typically 1000s of candidate 4-loops are found. B: Additional lines are predicted.  Pairs of new lines which form new 4-loops are added. C: Scaffold construction continues until no line is found, or the edge of the spectral range is reached.  A scaffold like this can only be assigned in a handful of ways. }
    \label{fig:construction}
\end{figure}

A \emph{4-loop} refers to a set of four transitions that form a closed path from one rotational state back to itself.  Conservation of energy constrains such frequencies to add and subtract up to zero, as in Figures \ref{fig:patterns}A and \ref{fig:patterns}B.  A randomly selected group of four lines is therefore highly unlikely to form a plausible 4-loop.  This constraint depends \emph{only} on conservation of energy, and so will robustly hold even for non-rigid rotor spectra.%  We note the distinction between 4-loops and closely spaced ``quartets'' in the literature commonly used to start an assignment; unlike quartets, 4-loop frequencies must sum to effectively zero and need not be closely spaced \cite{lesarri2011quartet}. 
The constraints on 4-loops are closely related to the established method of using ground state combination-differences to assign infrared or optical spectra \cite{sung2012extended,tellinghuisen2003combination}.  Convergence of 4-loops are routinely used to confirm experimental assignments \cite{matsushima2001frequency,lesarri2011quartet}.  

%From ref matsushima, for google terms: The number of combination loops composed of four transitions was rather small compared to that in the previous works for ground state [28–30], because many of the lines which are necessary to compose the frequency loops were not measured because of their weakness. A statistics of the loop defects of the combination loops are shown in Fig. 3 as a histogram.

A \emph{series} is a set of correlated lines of varying $J$.  Series can be $a$-type, $b$-type, or $c$-type. Figure \ref{fig:patterns}C shows a typical series in myrtenal (6,6-dimethylbicyclo[3.1.1]hept-3-ene-4-carbaldehyde, C$_{10}$H$_{14}$O).  The frequencies of a given series are nearly evenly spaced, and in particular are well fit by a quadratic polynomial, with fit residuals typically less than 1\%. $a$-type series with $J \gtrsim 5$ are particularly well fit by a quadratic polynomial. This regular spacing is exploited in Loomis-Wood plots \cite{lodyga2007advanced,winnewisser1989interactive}. As with 4-loops, a randomly selected group of lines is unlikely to form a plausible series.

The RAARR algorithm works by finding sets of four series, or ``scaffolds,'' that fit together following the specific structure shown in Figure \ref{fig:patterns}D.  Each scaffold contains several 4-loops and four series, and is thus highly constrained. While many types of 4-loops exist in a spectrum, {if the approximate selection rules of $\Delta k_a = 0,\pm1$ and $\Delta k_c = 0,\pm1$ are obeyed}, only 4-loops comprised of two $a$-type and two $b$-type R-branch transitions can be assembled in the specific, interwoven ``scaffold'' structure shown in Figure \ref{fig:patterns}D.  {If transitions with $\Delta k_c = \pm2$ are also considered, 4-loops comprised of two $a$-type and two $c$-type R-branch transitions can also be assembled in a similar structure.  While our algorithm attempts such assignments, the weak $\Delta k_c = \pm2$ $c$-type transitions in these scaffolds makes scaffolds of this form difficult to detect.} 

%XXX THINK ABOUT b-c type ``scaffolds'' is previous statement true?

  While the sum $S$ of the frequencies of a perfectly measured 4-loop should be strictly zero, measurement errors inevitably lead to small non-zero value for $S$.  We measured  $S$ for several hundred 4-loops in diverse molecules, and found $|S| < 50$ kHz more than 99\% of the time (see supplementary Figure 2).  We used 50 kHz as our acceptance window, and found that algorithm performance was not sensitive to small changes in this value.  We anticipate that this value would need to be reset depending on the spectrometer used.

%We focus on finding four series spaced $B + C$ apart, of the forms
%\begin{align*}
%    &\ket{J,K_{a0},J-K_{a0}} &\rightarrow \qquad \qquad \qquad &\ket{J+1,K_{a0},J-K_{a0}+1}\\ &\ket{J,K_{a0},J-K_{a0}} &\rightarrow \qquad \qquad \qquad &\ket{J+1,K_{a0}+1,J-K_{a0}}\\
%    &\ket{J,K_{a0}+1,J-K_{a0}-1} &\rightarrow \qquad \qquad \qquad &\ket{J+1,K_{a0}+1,J-K_{a0}}\\ &\ket{J,K_{a0}+1,J-K_{a0}-1} &\rightarrow \qquad \qquad \qquad &\ket{J+1,K_{a0},J-K_{a0}+1}
%\end{align*}
%with $K_{a0} = 0$, $1$, or $2$.

\subsection{Scaffold construction}
\label{section:variant1}
Scaffolds are built sequentially, from an initial 4-loop.  Candidate scaffolds which do not simultaneously fit constraints from four-loop convergence and series behavior are discarded as they are found.

\begin{enumerate}
    \item RAARR finds all plausible 4-loops containing at least one strong observed line. In a typical congested spectrum, this step finds between 500 and 5000 candidate 4-loops; the majority of these are coincidence, and do not correspond to a connected set of lines.  Others are legitimate 4-loops which cannot be extended into scaffolds; four-loops which contain Q-branch transitions fall into this category. See Figure \ref{fig:construction}A.
    
    \item The algorithm predicts two pairs of frequencies, each necessary to grow the scaffold in either direction.  At least one line from each pair is the next in a series, and each pair of lines combined with the existing scaffold completes a new 4-loop. See Figure \ref{fig:construction}B.
    
    \item  If a pair of lines close to the predicted frequencies and intensities is found, it is added to the scaffold.
    
    \item Scaffolds which contain series which are not well fit by a quadratic polynomial are discarded. Scaffolds comprised of lines that do not fit the intensity constraints described in Section \ref{intensesection} are also discarded. This discarding step is critical to efficiently trimming a potentially enormous list of candidate scaffolds.
    
    \item Steps 2-4 are repeated until additional lines are not found and/or the edge of the spectrum has been reached.  
    
    \item Each scaffold is assigned a score, or $p$-value, which estimates the chance that a structure as constrained as the scaffold in question or better would arise by chance. $p$-values as low as $10^{-50}$ are routinely found, indicating that the algorithm is confident before any attempt at assignment that the scaffold reflects a connected set of lines.
    
    \item The scaffolds with the lowest $p$-value are assigned using SPFIT, as described below. 
    
\end{enumerate}

The myrtenal spectrum shown in Figure \ref{fig:scaffoldfig} represents a relatively uncongested spectrum.  In this case, 1966 4-loops were considered.  These were rapidly ($\sim$20 seconds) winnowed down to 2 scaffolds, each comprised of about 20 spectral lines.  In more congested spectra, many thousand potential 4-loops are found, and the algorithm typically takes several minutes to find a scaffold.  %Although the algorithm can be set to keep searching ad infinitum, we have not yet seen a case where it finds a scaffold after five minutes but not before.
%n spectra   How many four loops typically found? 2 level ladders?  3 level ladders? final p-value? final assigned lines? Walk through it for one case, maybe myrtenal

\subsection{Intensity information} \label{intensesection}

  In many spectrometers, including our own, line intensities can vary significantly, and so the algorithm must be robust against this variation.  Line intensities are used in RAARR in the following ways: 1) the initial 4-loop is assumed to have at least one line in the top 40 most intense lines, 2) variation of line intensities in a given \emph{series} can be no more than a factor of 10, and 3) variation of line intensities in a given \emph{scaffold} can be no more than a factor of 100.  
  These constraints allow the common situation where $|\mu_a|$ and $|\mu_b|$, which we cannot determine prior to assignment, differ significantly.  We have found that algorithm performance is not highly sensitive to adjustments in these parameters.  {The third constraint effectively prevents the algorithm from finding analogous scaffolds comprised of $a$-type and $c$-type transitions, which necessarily contain weak $c$-type lines with $\Delta_{kc} = \pm2$}. % We attempted to find these scaffolds, and thus assign spectra without $b$-type transitions, by loosening this constraint, but could not find conditions which would efficiently find such scaffolds, which exhibit high variation in line intensities.}

\subsection{Assignment of scaffolds}
  To assign a scaffold, we need to determine which of its four series are $a$-type and which are $b$-type, and to guess the upper and lower states of each transition in the scaffold.  We then call Pickett's SPFIT, which, if the fit converges, will return three rotational constants and the RMS error on the scaffold lines.  Lastly, we use Pickett's SPCAT to generate a theoretical spectrum from the constants, which we compare against the experimental spectrum.
  
  We can unambiguously distinguish $a$- and $b$-type series by the signs of the frequencies in the following equation, which each 4-loop must satisfy,
  \begin{equation} \label{eq:signs}
       f_1 \pm f_2 \pm f_3 \pm f_4 = 0
  \end{equation}
  Consider 4-loops containing exactly two lines from the same series.  If these two same-series lines have the same sign in Equation \ref{eq:signs}, they must be $a$-type transitions (between states of the same $K_a$, as in Figure \ref{fig:patterns}A); else, they must be $b$-type (between states with $K_a$ differing by one, as in Figure \ref{fig:patterns}B).

 Of the two $a$-type series, one connects a series of states $\ket{{J_n}_{\,k_0,\,J_n-k_0}}$, and the other a series of states $\ket{{J_n}_{\,k_0+1,\,J_n-k_0}}$. The first of these series is shown as the blue, left hand series in Figures \ref{fig:patterns}D and \ref{fig:scaffoldfig}B.  This series can be identified prior to assignment by comparing the energies of the left hand and right hand states. We use $k_0$ to denote the minimum $K_a$ found in the scaffold.  The state $\ket{{J_n}_{\,k_0,\,J_n-k_0}}$ has strictly lower energy than the corresponding state with $\ket{{J_n}_{\,k_0+1,\,J_n-k_0}}$ in both prolate and oblate asymmetric tops.
 
    With the series identified, the assignment of each state in the entire scaffold is determined by $J$ and $K_a$ of the lowest state. $J$ can be estimated from the $\ket{{J_n}_{\,k_0+1,\,J_n-k_0}}$ $a$-type series (black lines), by noting that $J \approx J_{guess} = \nu_1/(\nu_2-\nu_1)$, where $\nu_1$ is the frequency of the lowest $a$-type transition in this series, and $\nu_2$ is the frequency of the second lowest transition (see for example the analytical approximations described in Cooke et al. \cite{cooke2012decoding}).  Although we find scaffolds in most cases with $K_a$ values of 0, 1, or 2, we try assignments up to $K_a = 4$.  We find the most robust performance is achieved by trying a range of $J = [J_{guess}-2..J_{guess}+2]$ and $K_a = [0..4]$ for the lowest state in the scaffold.
    
    {Scaffolds of the type shown in figure \ref{fig:patterns} can also be formed via a combination of $a$-type and $c$-type transitions if $c$-type transitions with $\Delta k_a = \pm2$ are considered.  For these assignments the $b$-type transitions shown in red and purple in figure \ref{fig:patterns} are replaced by $c$-type transitions, and the states $\ket{{J_n}_{\,k_0+1,\,J_n-k_0}}$ are replaced with $\ket{{J_n}_{\,k_0+1,\,J_n-k_0-1}}$ . No such scaffolds were found for the molecules shown in table \ref{table:successful_fits}. The $c$-type transitions with $\Delta k_a = \pm2$, although detectable in our spectra, typically had measured intensities of less than 1\% of our strong $a$-type lines, and scaffolds which included these lines were disallowed as discussed in section \ref{intensesection}.}
\begin{table}[h]
\centering
\caption{A scaffold is comprised of 4 series: 2 of $a$-type transitions, and 2 of $b$-type transitions.  In our figures, we represent $a$-type series with vertical blue or black lines and $b$-type series with diagonal red or purple lines.}
\label{table:K_c_table}
\begin{tabular}{|l|l|l|l|}
\hline
Series & Color & Upper state $K_c$ & Lower state $K_c$ \\
\hline
a-type & Blue & $J - k_0$ & $J - k_0$       \\
b-type & Red & $J - k_0$ & $J - k_0 + 1$       \\
b-type & Purple & $J - k_0 + 1$ & $J - k_0$       \\
a-type & Black & $J - k_0 + 1$ & $J - k_0 + 1$       \\
\hline
\end{tabular}
\end{table}

\section{Performance}

We ran our algorithm on three qualitatively distinct sets of spectra: simulated spectra of ``fake'' rigid rotor molecules, experimental spectra of molecules that have been previously assigned, and experimental spectra of previously unassigned molecules.

\subsection{Performance on simulated spectra}
  A large number ($N \approx 10000)$ of ``fake'' molecular spectra were generated.  To generate these spectra, $[A,B,C]$ and centrifugal distortion constants were chosen within varying ranges, and the rotational spectrum was calculated.  The ideal spectra produced were then distorted to simulate realistic instrumentation errors: lines were shifted in frequency by random amounts within $\pm 5$ kHz, and noise was added.  In addition, lines were broadened and the heights of the lines were randomized rather aggressively by a factor of $0.3-3$, reflecting our experience that experimental line strengths are unreliable.  The broadening and resampling of the lines leads to the blending of nearby lines, as happens experimentally. A significant number of random lines were added to the spectra, to simulate lines corresponding to excited states, isotopologs, conformers, and chemical contaminants.  Spectra were cut off outside an adjustable range, reflecting the finite range of the simulated instrument. It is noteworthy that despite these significant efforts to ``spoil'' our simulated spectra, RAARR still performed significantly better on the simulated spectra than on experimental spectra.
  
  The algorithm correctly assigned these simulated spectra in more than 99\% of the cases which fit the following three criteria:
  \label{sec:crtiteria} 
  
  \begin{enumerate}
      \item  A scaffold connecting levels of at least 4 distinct $J$ values can be found within the simulated range of the instrument.  This constraint is essentially guaranteed for molecules with $B+C \lesssim (f_{max} - f_{min})/4$ and $A \lesssim (f_{max} - f_{min})/2$. Very small molecules, with large $A$, $B$, and $C$, typically fail this criteria
      
      \item The molecule is not a symmetric or nearly symmetric top. The highly repetitive spectra of nearly symmetric tops leads to many false-positive scaffolds, confusing the algorithm. In simulated spectra this only led to failure for molecules with Ray's asymmetry parameter $\kappa \lesssim -0.995$, or $B-C \lesssim $ 2 MHz
      
      \item Both $a$-type and $b$-type transitions are visible in the spectrum
      
  \end{enumerate}
  
\begin{figure}[ht!] %made by C:Dropbox\squareassign\repeat42
\includegraphics[width=5in]{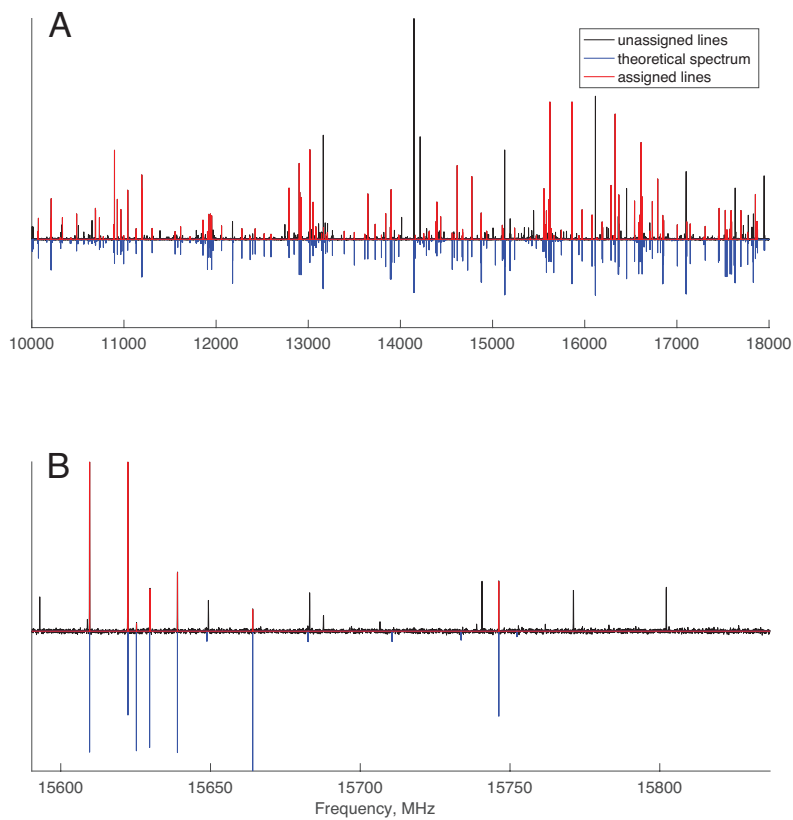}%fake_fig.fig,fake_fig.eps}
\caption{A successful fit on simulated data.  A: The complete simulated spectrum, from 10 GHz to 18 GHz.  Noise, distortions, and false lines have been added as described in the text.  B: a magnified portion showing a few of the many successfully assigned lines.  In this case the assignment was $A$ = 1220.3 MHz, $B$ = 674.0 MHz, and $C$ = 492.3 MHz; the spectrum was calculated from $A$ = 1220.4 MHz, $B$ = 674.0 MHz, and $C$ = 492.3 MHz.
\label{fakefig}}
\end{figure}

\begin{figure}[ht!] %made by C:Dropbox\squareassign\makemyrtenal
\includegraphics[width=6in]{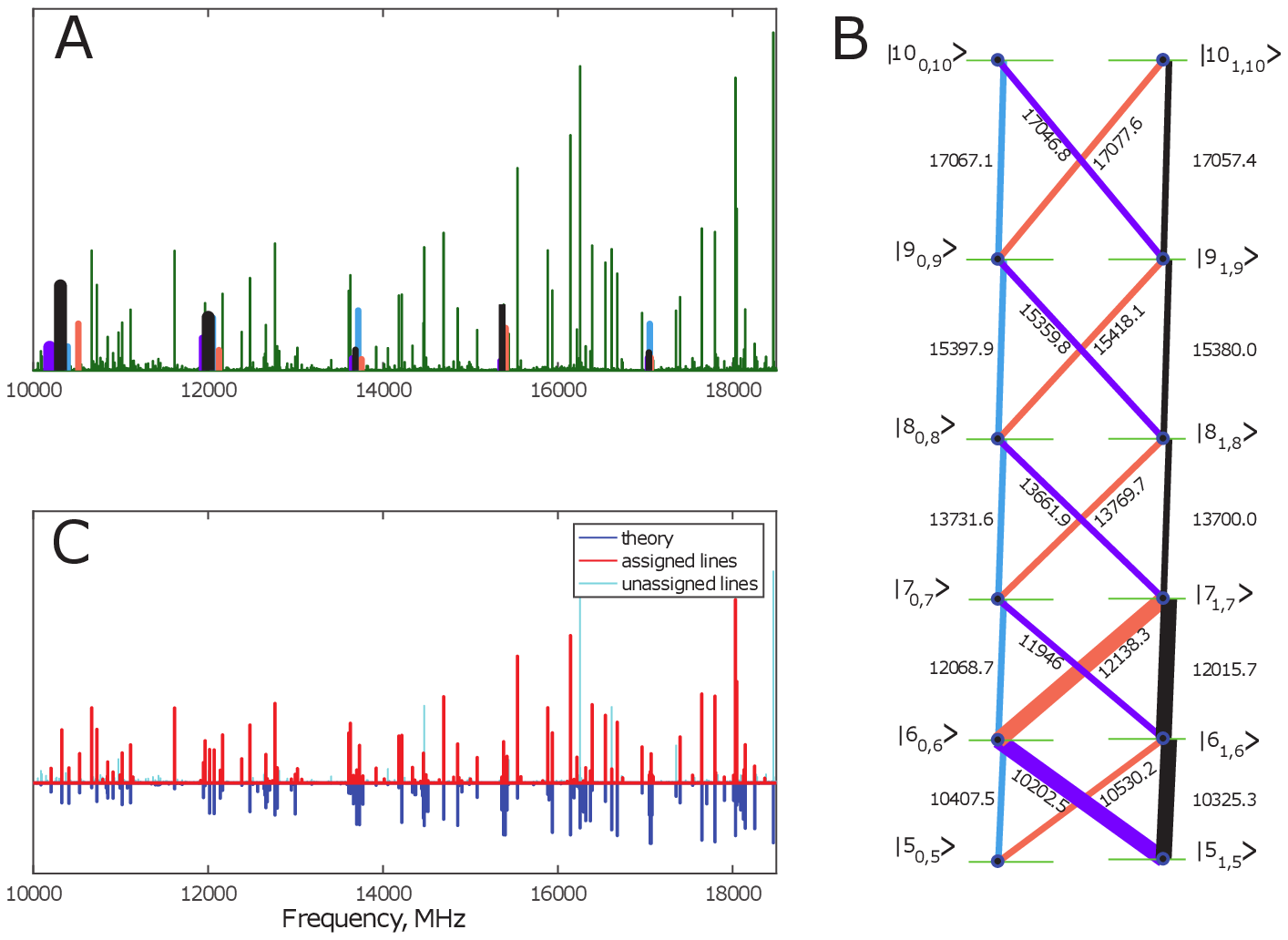}%ionapparatus.eps}
\caption{A successful fit on the molecule myrtenal. A: The unassigned experimental spectrum (green), with the lines used for the scaffold marked in color.  The original 4-loop at 10325.3 MHz, 12015.7 MHz, 12138.3 MHz, and 10202.5 MHz is marked in bold. B: The scaffold found by extending this 4-loop.  Colors correspond to those shown in A. C: The assignment found from this scaffold.  The lower blue spectrum is the predicted spectrum from the automatically assigned rotational constants of A = 1666.42 MHz, B = 962.37 MHz, and C = 836.90 MHz.  The red lines show corresponding assigned experimental lines.  This automatic fit takes about 40 seconds to calculate. %A second conformer (A = 2329.09 MHz, B = 620.77 MHz, C = 562.41 MHz) was also found
\label{fig:scaffoldfig}}
\end{figure}

  %Although RAARR does not explicitly include centrifugal distortion, it can tolerate magnitudes of centrifugal distortion typical to most asymmetric top molecules for low temperature spectra.
  The values found by the algorithm for $A$, $B$, and $C$ differed very slightly from the original values; this is expected because of the intentional distortions in the simulated spectra.  
  Typical output from running RAARR on a simulated spectrum is shown in Figure \ref{fakefig}.
  
  %  The algorithm succeeded on more than 99\% of the molecules which fit these three criteria.

\subsection{Performance on experimental spectra}
 Table \ref{table:successful_fits} shows results of running RAARR on a variety of experimental spectra. 15 molecules in all were assigned, including 5 new species; one species (myrtenal) had been previously assigned but we were unaware of this at the time of our assignment.  RAARR agreed with previous assignments in all cases where a previous assignment had been made.
 
 We also applied RAARR to 5 other species (heptaldehyde, rosemary oil, butanol, benzyl alcohol, and linalool oxide) without success. % We believe that either these molecules are too small to have enough lines within our experimental range, or that $a$-type or $b$-type transitions are absent in these spectra. XXX
 
\begin{table}[!ht]
    \centering
    \begin{tabular}{|c|c|c|c|}
        \hline
        Molecule & Previous assignment & \multicolumn{2}{|c|}{Our assignments}\\
        \cline{3-4}
         & & variant 1 & variant 2\\
        \hline
        \hline
        $\alpha$-pinene&1936.6, 1228.6, 1127.0 \cite{alphapineneconstants}&1936.6, 1228.6, 1127.0&1936.6, 1228.6, 1127.0\\
        \hline
        $\alpha$-terpineol&N/A&\begin{tabular}[c]{@{}l@{}}2329.3, 619.0, 560.4\\ 2329.1, 620.8, 562.4\end{tabular}&2329.1, 620.8, 562.4\\
        \hline
        benzaldehyde&5234.3, 1564.2, 1204.7 \cite{benzaldehydeconstants}&\begin{tabular}[c]{@{}l@{}}5234.4, 1564.3, 1204.7\\5229.6, 1564.5, 1205.4\end{tabular}&5234.4, 1564.3, 1204.7\\
        \hline
        benzyl-$\alpha$-D$_1$ alcohol&N/A&\begin{tabular}[c]{@{}l@{}}4679.2, 1443.0, 1169.2\\ 4667.7, 1452.1, 1163.6\end{tabular}&spectrum too short\\
        \hline
        $\beta$-pinene&1901.9, 1293.7, 1150.8 \cite{betapineneconstants}&1901.9, 1293.7, 1150.8&1901.9, 1293.7, 1150.8\\
        \hline
        cinnamaldehyde&4866.4, 579.1, 517.8 \cite{cinnamaldehydeconstants}&4866.4, 579.1, 517.8&4866.4, 579.1, 517.8\\
        \hline
        ethylguiacol&N/A&1640.6, 802.5, 567.8&1640.6, 802.5, 567.8\\
        \hline
        limonene (from lemon oil)&3058.0, 717.0, 679.3 \cite{moreno2013conformational}&3058.0, 717.0, 679.3&3058.0, 717.0, 679.3\\
        \hline
        limonene (from orange extract)&3058.0, 717.0, 679.3 \cite{moreno2013conformational}&\begin{tabular}[c]{@{}l@{}}3058.0, 717.1, 679.33\\3053.9, 742.2, 646.7\\3040.8, 745.7, 643.8\end{tabular}&3041.2, 745.7, 643.7\\
        \hline
        myrtenal&1666.3, 962.3, 836.9 \cite{myrtenalconstants}&1666.4, 962.4, 836.9&1666.7, 962.4, 836.9\\
        \hline
        myrtenol&1589.4, 971.9, 827.7 \cite{sedo1918cp}&\begin{tabular}[c]{@{}l@{}}1589.4, 971.92, 827.7\\1701.6, 871.7, 802.2\end{tabular}&1702.1, 871.7, 802.1\\
        \hline
        menthone&1953.4, 694.5, 586.6 \cite{menthoneconstants}&\begin{tabular}[c]{@{}l@{}}1953.7, 694.5, 586.6\\2022.0, 693.5, 562.1\end{tabular}&1953.5, 694.5, 586.6\\
        \hline
        m-anisaldehyde&N/A&\begin{tabular}[c]{@{}l@{}}1939.9, 1087.9, 700.3\\2528.1, 894.1, 663.5\\2619.3, 880.9, 662.2\end{tabular}&1940.0, 1088.0, 700.3\\
        \hline
        nopinone&1923.1, 1297.6, 1164.0 \cite{betapineneconstants}&1923.3, 1297.6, 1164.0&1923.2, 1297.6, 1164.0\\
        \hline
        p-anisaldehyde&4160.7, 717.8, 614.9 \cite{bohn1992rotational}&\begin{tabular}[c]{@{}l@{}}4661.9, 698.4, 610.1\\4160.7, 717.8, 614.9\end{tabular}&4160.7, 717.8, 614.9\\
        \hline
        phlorol&N/A&\begin{tabular}[c]{@{}l@{}}1906.0, 463.1, 443.2\\ 1881.6,      457.5,      438.7\end{tabular}&1906.0, 463.1, 443.2\\
        \hline
    \end{tabular}\\
    \caption{Experimental spectra assigned with RAARR.  We have two variants of the RAARR algorithm which differ in how they build the scaffold.  See the main body of the paper, in particular Section \ref{section:variant1}, for details on variant 1; supplementary Section 1.2 for details on variant 2; and supplementary Section 1.1 for runtimes.  Variant 1 has conformer search implemented, while variant 2 outputs only the first set of rotational constants for which Pickett's SPCAT converges. Variant 2 is significantly faster, but slightly less robust on congested simulated spectra. Only one conformer is listed from previous assignments.}
    \label{table:successful_fits}
\end{table}

\subsection{Performance on mixtures}
The algorithm was tested on several mixtures, and was able to assign $A$, $B$, and $C$ independently to each of several mixture components.  The most natural mixture is conformers from a single compound; as seen in Table \ref{table:successful_fits}, the RAARR algorithm successfully found 2 or more conformers in 9 separate cases.  In the case of limonene, the fit was found independently both from a sample of lemon oil and a sample of orange extract.  In several other cases, fits were found despite the spectra being heavily contaminated with ethanol.  While a known contaminant such as ethanol could be removed from the spectrum, this was not done here.  Ethanol is both non-rigid and too small to satisfy $B+C \lesssim (f_{max} - f_{min})/4$, as described in Section \ref{sec:crtiteria}.

While the algorithm performs well on mixture components which are substantially represented in the sample, at present it performs poorly on trace contaminants.  Our sample of lemon oil contained a visible $\alpha$-pinene spectrum at the few percent level, but the RAARR algorithm did not find this; in addition, $^{13}$C isotopomers are in general too weak to be independently assigned.  The current algorithm cannot find these trace components because the experimental spectra are highly congested when such weak lines are considered, and the number of 4-loops and scaffolds becomes untenable. We are optimistic that future extensions to the algorithm, which take advantage of the fact that the parent isotopomer has very similar $A$, $B$, and $C$ and centrifugal distortion constants to $^{13}$C isotopomers, will be able to find scaffolds of $^{13}$C isotopomers even amid such congested spectra.  Automated assignment of all singly-substituted isotopomers of all visible conformers is a prerequisite to fully automated context-free structural determination, which is a long-standing goal of the microwave spectroscopy community.

\subsection{Limitations}
The current algorithm only works for asymmetric rigid rotor molecules with visible $a$-type and $b$-type lines.  We have seen cases where although the algorithm is ultimately unsuccessful,  it nevertheless finds highly constrained scaffolds, which are strongly indicated to comprise connected sets of lines.  These scaffolds then resist assignment to rigid rotors.  It is our belief that these molecules are non-rigid, for example exhibiting internal tunneling behavior.  Running RAARR on spectra of \emph{known} non-rigid molecules, such as benzyl alcohol, similarly found scaffolds but no assignment. Extensions to our algorithm which allow for fitting Hamiltonians with internal barriers would therefore likely assign these molecules as well.

To date, the algorithm has only been tested on low temperature spectra, such as those produced in buffer gas cooled or pulsed jet spectrometers.  While the basic approach of finding scaffolds within complex spectra should apply at higher temperatures, the algorithm would likely have to be adapted; in particular, the current approach of fitting spectra first without centrifugal distortion and then adding these terms is likely inapplicable to spectra containing high $J$ lines.  Larger species are likely intractable at higher temperatures, simply because the spectra are congested to the point where individual lines cannot be resolved.

Our restriction to rigid rotor molecules additionally precludes molecules with visible hyperfine structure.  In the context of organic compounds, this limitation is most notable in that the current algorithm gets confused by the hyperfine structure arising from $^{14}$N nuclear electric quadrupole splittings in compounds containing nitrogen.  We anticipate that future versions of RAARR will be able to assign spectra of nitrogen-containing compounds as well, although this will become substantially more difficult for compounds with several nitrogens. 

We are developing further algorithms which assign spectra based on combinations of $a$-type and $c$-type lines, combinations of $b$-type and $c$-type lines, and multiple series of $a$-type lines alone.  %In principal, the algorithm presented here could detect scaffolds comprised of $a$-type and $c$-type lines, but the weak $c-type$ transitions with $\Delta k_a} = \pm 2$ in such spectra are typically rejected by the scaffold-building alXXX
In the case of using $a$-type lines alone, longer series are required; a reasonable constraint would be $B+C \lesssim (f_{max} - f_{min})/6$, rather than $B+C \lesssim (f_{max} - f_{min})/4$ as is required here.  Each member of the resulting family of algorithms will be applied to an unassigned spectrum, substantially increasing the probability of a successful assignment.
The performance of these algorithms will be described in a future publication.
\section{Conclusion}

We have demonstrated a new algorithm, RAARR, which can rapidly assign experimental rigid rotor asymmetric top spectra, including mixtures of conformers and mixtures of different compounds.  The algorithm finds assignments from a broadband spectrum for diverse types of molecules, primarily limited by the constraint that the molecules exhibit visible $a$-type and $b$-type lines.  The algorithm is shown experimentally to be capable of assigning mixtures.

\section{Description of supplementary material}
The runtimes of RAARR (at the time of publication) on each molecule in this paper are listed.  We also describe a second variant of the RAARR algorithm that searches for scaffolds more efficiently for typical spectra via exploiting overlapping frequencies in scaffold 4-loops.  We present a histogram of our experimental convergence of 4-loop sums to justify the bounds on it discussed in Section \ref{sec:pattern}.

\section{Acknowledgements}
We thank Crist\'{o}bal P\'{e}rez at the Max-Planck Institute for Structure and Dynamics of Matter for many helpful discussions, and assigning spectra we couldn't. We thank the Packard Foundation and the National Science Foundation for financial support (NSF grant DBI-1832846).

\clearpage
\bibliographystyle{ieeetr}
\bibliography{QC.bib}

\begin{thebibliography}{10}

\bibitem{cooke2012decoding}
S.~Cooke and P.~Ohring, ``Decoding pure rotational molecular spectra for
  asymmetric molecules,'' {\em Journal of Spectroscopy}, vol.~2013, 2012.

\bibitem{hageman2000direct}
J.~Hageman, R.~Wehrens, R.~de~Gelder, W.~Leo~Meerts, and L.~Buydens, ``Direct
  determination of molecular constants from rovibronic spectra with genetic
  algorithms,'' {\em The Journal of Chemical Physics}, vol.~113, no.~18,
  pp.~7955--7962, 2000.

\bibitem{seifert2015autofit}
N.~A. Seifert, I.~A. Finneran, C.~Perez, D.~P. Zaleski, J.~L. Neill, A.~L.
  Steber, R.~D. Suenram, A.~Lesarri, S.~T. Shipman, and B.~H. Pate,
  ``{AUTOFIT}, an automated fitting tool for broadband rotational spectra, and
  applications to 1-hexanal,'' {\em Journal of Molecular Spectroscopy},
  vol.~312, pp.~13--21, 2015.

\bibitem{shipman2019autofit}
E.~J. Riffe, S.~T. Shipman, S.~A. Gaster, C.~M. Funderburk, and G.~G. Brown,
  ``Rotational spectrum of eugenol as analyzed with double resonance and
  grid-based autofit,'' {\em The Journal of Physical Chemistry A}, vol.~123,
  no.~5, pp.~1091--1099, 2019.
\newblock PMID: 30640470.

\bibitem{martin2016automated}
M.-A. Martin-Drumel, M.~C. McCarthy, D.~Patterson, B.~A. McGuire, and K.~N.
  Crabtree, ``Automated microwave double resonance spectroscopy: A tool to
  identify and characterize chemical compounds,'' {\em The Journal of Chemical
  Physics}, vol.~144, no.~12, p.~124202, 2016.

\bibitem{fritz2018conformer}
S.~M. Fritz, A.~Hernandez-Castillo, C.~Abeysekera, B.~M. Hays, and T.~S. Zwier,
  ``Conformer-specific microwave spectroscopy of 3-phenylpropionitrile by
  strong field coherence breaking,'' {\em Journal of Molecular Spectroscopy},
  vol.~349, pp.~10--16, 2018.

\bibitem{zaleski2018automated}
D.~P. Zaleski and K.~Prozument, ``Automated assignment of rotational spectra
  using artificial neural networks,'' {\em The Journal of Chemical Physics},
  vol.~149, no.~10, p.~104106, 2018.

\bibitem{western2017automatic}
C.~M. Western and B.~E. Billinghurst, ``Automatic assignment and fitting of
  spectra with {PGOPHER},'' {\em Physical Chemistry Chemical Physics}, vol.~19,
  no.~16, pp.~10222--10226, 2017.

\bibitem{leo2006application}
W.~Leo~Meerts and M.~Schmitt, ``Application of genetic algorithms in automated
  assignments of high-resolution spectra,'' {\em International Reviews in
  Physical Chemistry}, vol.~25, no.~3, pp.~353--406, 2006.

\bibitem{schmitt2004determination}
M.~Schmitt, C.~Ratzer, K.~Kleinermanns, and W.~Leo~Meerts, ``Determination of
  the structure of 7-azaindole in the electronic ground and excited state using
  high-resolution ultraviolet spectroscopy and an automated assignment based on
  a genetic algorithm,'' {\em Molecular Physics}, vol.~102, no.~14-15,
  pp.~1605--1614, 2004.

\bibitem{moseley1999automated}
H.~N. Moseley and G.~T. Montelione, ``Automated analysis of nmr assignments and
  structures for proteins,'' {\em Current opinion in structural biology},
  vol.~9, no.~5, pp.~635--642, 1999.

\bibitem{evangelidis2018automated}
T.~Evangelidis, S.~Nerli, J.~Nov{\'a}{\v{c}}ek, A.~E. Brereton, P.~A. Karplus,
  R.~R. Dotas, V.~Venditti, N.~G. Sgourakis, and K.~Tripsianes, ``Automated nmr
  resonance assignments and structure determination using a minimal set of 4d
  spectra,'' {\em Nature communications}, vol.~9, no.~1, p.~384, 2018.

\bibitem{PGOPHER}
C.~M. Western, {\em PGOPHER, a Program for Simulating Rotational Structure}.
\newblock University of Bristol.

\bibitem{pickett}
H.~M. Pickett, {\em THE {SPFIT} PROGRAM SET. These programs are copyrighted,
  but made freely available as a service to the spectroscopic community. A
  manuscript detailing some of the uses of the software is: B. J. Drouin,
  "Practical Uses of SPFIT," J. Molec. Spectroscopy 340, 1-15, 2017. The
  program suite is located at \verb|https://spec.jpl.nasa.gov/|}.
\newblock Ohio State University, 2001.

\bibitem{GordyCook}
W.~Gordy and R.~L. Cook, {\em Microwave molecular spectra}.
\newblock Wiley New York, 3rd ed.~ed., 1984.

\bibitem{sung2012extended}
K.~Sung, L.~R. Brown, X.~Huang, D.~W. Schwenke, T.~J. Lee, S.~L. Coy, and K.~K.
  Lehmann, ``Extended line positions, intensities, empirical lower state
  energies and quantum assignments of {NH$_3$} from 6300 to 7000 cm$^{-1}$,''
  {\em Journal of Quantitative Spectroscopy and Radiative Transfer}, vol.~113,
  no.~11, pp.~1066--1083, 2012.

\bibitem{tellinghuisen2003combination}
J.~Tellinghuisen, ``Combination differences: Victim of false charges?,'' {\em
  Journal of Molecular Spectroscopy}, vol.~221, no.~2, pp.~244--249, 2003.

\bibitem{matsushima2001frequency}
F.~Matsushima, M.~Matsunaga, G.-Y. Qian, Y.~Ohtaki, R.-L. Wang, and K.~Takagi,
  ``Frequency measurement of pure rotational transitions of {D$_2$O} from 0.5
  to 5 {THz},'' {\em Journal of Molecular Spectroscopy}, vol.~206, no.~1,
  pp.~41--46, 2001.

\bibitem{lesarri2011quartet}
A.~Lesarri, A.~Vega-Toribio, R.~D. Suenram, D.~J. Brugh, D.~Nori-Shargh, J.~E.
  Boggs, and J.-U. Grabow, ``Structural evidence of anomeric effects in the
  anesthetic isoflurane,'' {\em Physical Chemistry Chemical Physics}, vol.~13,
  pp.~6610--6618, 2011.

\bibitem{lodyga2007advanced}
W.~Lodyga, M.~Kreglewski, P.~Pracna, and {\v{S}}.~Urban, ``Advanced graphical
  software for assignments of transitions in rovibrational spectra,'' {\em
  Journal of Molecular Spectroscopy}, vol.~243, no.~2, pp.~182--188, 2007.

\bibitem{winnewisser1989interactive}
B.~P. Winnewisser, J.~Reinst{\"a}dtler, K.~M. Yamada, and J.~Behrend,
  ``Interactive {Loomis-Wood} assignment programs,'' {\em Journal of Molecular
  Spectroscopy}, vol.~136, no.~1, pp.~12--16, 1989.

\bibitem{alphapineneconstants}
E.~M. Neeman, J.-R. Avil\'es-Moreno, and T.~R. Huet, ``The gas phase structure
  of $\alpha$-pinene, a main biogenic volatile organic compound,'' {\em The
  Journal of Chemical Physics}, vol.~147, p.~214305, 2017.

\bibitem{benzaldehydeconstants}
R.~K. Kakar, E.~A. Rinehart, C.~R. Quade, and T.~Kojima, ``Microwave spectrum
  of benzaldehyde,'' {\em The Journal of Chemical Physics}, vol.~52, no.~7,
  pp.~3803 -- 3813, 1970.

\bibitem{betapineneconstants}
E.~M. Neeman, J.-R. Avil\'es-Moreno, and T.~R. Huet, ``The quasi-unchanged
  gas-phase molecular structures of the atmospheric aerosol precursor
  $\beta$-pinene and its oxidation product nopinone,'' {\em Physical Chemistry
  Chemical Physics}, vol.~19, pp.~13819 -- 13827, 2017.

\bibitem{cinnamaldehydeconstants}
S.~Zinn, T.~Betz, C.~Medcraft, and M.~Schnell, ``Structure determination of
  trans-cinnamaldehyde by broadband microwave spectroscopy,'' {\em Physical
  Chemistry Chemical Physics}, vol.~17, pp.~16080 -- 16085, 2015.

\bibitem{moreno2013conformational}
J.~R.~A. Moreno, T.~R. Huet, and J.~J.~L. Gonz{\'a}lez, ``Conformational
  relaxation of s-(+)-carvone and r-(+)-limonene studied by microwave fourier
  transform spectroscopy and quantum chemical calculations,'' {\em Structural
  Chemistry}, vol.~24, no.~4, pp.~1163--1170, 2013.

\bibitem{myrtenalconstants}
M.~Chrayteh, P.~Dr\'ean, and T.~R. Huet, ``Structure determination of myrtenal
  by microwave spectroscopy and quantum chemical calculations,'' {\em Journal
  of Molecular Spectroscopy}, vol.~336, pp.~22 -- 28, 2017.

\bibitem{sedo1918cp}
G.~Sedo {\em et~al.}, ``{CP-FTMW} spectroscopy of the low energy conformers of
  two chiral alcohols: myrtenol and nopol,'' International Symposium on
  Molecular Spectroscopy, 2018.

\bibitem{menthoneconstants}
D.~Schmitz, V.~A. Shubert, T.~Betz, and M.~Schnell, ``Exploring the
  conformational landscape of menthol, menthone, and isomenthone: a microwave
  study,'' {\em Frontiers in Chemistry}, vol.~3, 2015.

\bibitem{bohn1992rotational}
R.~K. Bohn, M.~S. Farag, C.~M. Ott, J.~Radhakrishnan, S.~A. Sorenson, and N.~S.
  True, ``{Rotational spectra of p-anisaldehyde. Assignment of the planar
  conformers and observation of torsionally excited states},'' {\em Journal of
  Molecular Structure}, vol.~268, no.~1-3, pp.~107--121, 1992.

\end{thebibliography}
\end{document}